\documentclass[aps,pre,twocolumn,showpacs,floatfix,superscriptaddress]{revtex4}
\usepackage{graphicx}
\usepackage{epsfig}
\usepackage{verbatim} 

\begin{document}

\title{Avalanches, branching ratios, and clustering of attractors
 in Random Boolean Networks and in the segment polarity network of \emph{Drosophila}}

\author{Andrew Berdahl}
\affiliation{Complexity Science Group, Department of Physics and Astronomy,
University of Calgary, Calgary, Alberta, Canada, T2N 1N4}

\author{Amer Shreim}
\affiliation{Complexity Science Group, Department of Physics and Astronomy,
University of Calgary, Calgary, Alberta, Canada, T2N 1N4}

\author{Vishal Sood}
\affiliation{Complexity Science Group, Department of Physics and Astronomy,
University of Calgary, Calgary, Alberta, Canada, T2N 1N4}

\author{J\"{o}rn Davidsen} 
\affiliation{Complexity Science Group, Department of Physics and Astronomy,
University of Calgary, Calgary, Alberta, Canada, T2N 1N4}

\author{Maya Paczuski}
\affiliation{Complexity Science Group, Department of Physics and Astronomy,
University of Calgary, Calgary, Alberta, Canada, T2N 1N4}

\date{\today}

\begin{abstract}
  We discuss basic features of emergent complexity in dynamical systems far
  from equilibrium by focusing on the network structure of their state space.
  We start by measuring the distributions of avalanche and transient times in
  Random Boolean Networks (RBNs) and in the \emph{Drosophila} polarity network by
  exact enumeration.  A transient time is the duration of the transient from a
  starting state to an attractor. An avalanche is a special transient which
  starts as single Boolean element perturbation of an attractor state.
  Significant differences at short times between the avalanche and the
  transient times for RBNs with small connectivity $K$ -- compared to the
  number of elements $N$ -- indicate that attractors tend to cluster in
  configuration space. In addition, one bit flip has a non-negligible chance
  to put an attractor state directly onto another attractor. This clustering
  is also present in the segment polarity gene network of \emph{Drosophila
    melanogaster}, suggesting that this may be a robust feature of biological
  regulatory networks.  We also define and measure a branching ratio for the
  state space networks and find evidence for a new time scale that diverges
  roughly linearly with $N$ for $2\leq K \ll N$.  Analytic arguments show that
  this time scale does not appear in the random map nor can the random map
  exhibit clustering of attractors. We further show that for $K=2$ the
  branching ratio exhibits the largest variation with distance from the
  attractor compared to other values of $K$ and that the avalanche durations
  exhibit no characteristic scale within our statistical resolution.  Hence,
  we propose that the branching ratio and the avalanche duration are new
  indicators for scale-free behavior that may or may not be found
  simultaneously with other indicators of emergent complexity in extended,
  deterministic dynamical systems.

\end{abstract}

\maketitle

\section{Introduction}

Random Boolean Networks (RBNs)~\cite{Kauffman1969} are elementary models for
signaling processes such as genetic regulation -- where a binary state based
on Boolean logic~\cite{drossel2007rbn,bornholdt05,aldanaRBN2003} encapsulates
local gene expression.  The dichotomy between their easy construction and
their emergent complex behavior has motivated researchers in diverse fields
including the neurological~\cite{wang1990oac},
computational~\cite{lynch1995tcr}, evolutionary~\cite{PaczuskiRBN} and
physical~\cite{baillie1994dqg,DerridaFlyvbjerg} sciences to use these or
related models as test beds for ideas about self-organization.

The most well known fact about RBNs is that they exhibit three distinct phases
in a statistical ensemble obtained by averaging over random realizations:
chaotic, frozen and critical, depending on the connectivity $K$ of the Boolean
elements. Derrida and Pomeau~\cite{DerridaPomeau} used an annealed
approximation to prove that $K=2$ is a critical ensemble in between ordered
and chaotic regimes.  According to their analysis, the distinct phases
correspond to different patterns of growth for the Hamming distance between
two nearly identical initial states.  For $K>2$ the distance on average grows
exponentially. For $K<2$ the distance on average decays exponentially. For
$K=2$ the ensemble of random realizations is critical - the distance is
dominated by fluctuations. The existence of this critical phase has been used
to argue that many natural systems, including life itself, function at a
so-called ``edge of chaos''~\cite{OriginOrder,bak,BakSandPile,PerBakGoL,
  packard1988ate}.

However, different measures of criticality, which appear simultaneously in the
$K=2$ ensemble of RBNs, may not point to the same critical behavior when
applied to real signaling networks.  For instance, there may be many distinct
``edges of chaos'' (according to different definitions of dynamical
criticality) that can appear in networks that are not members of a statistical
ensemble, but are, instead, organized by natural selection or by other forces.
More to the point, we believe that one should not conflate all measures of
criticality as being necessarily equivalent in treating complex systems far
from equilibrium since there is no reason \emph{a priori} that they should
probe intrinsically related dynamical fluctuations.  Hence, weq explore
several alternative probes of criticality in RBNs that may be more empirical
or, indeed, more useful in revealing the forces and constraints that shape
natural or man-made regulatory or signaling processes.

To this end, here we develop new methods of complex network analysis to
analyze ensembles of state space networks (SSNs) for RBNs and compare their
measurable observables to that of the segment polarity network of
\emph{Drosophila melanogaster}~\cite{albert2003dros} where possible.  The
aspects of these SSNs that have received most attention so far include the
probability distribution of attractor lengths and basins of attraction
sizes~\cite{greil07,Drossel2005PRE,samuelsson2003RBN}. In the context of
regulatory networks, attractors are thought to correspond to distinct cellular
states~\cite{OriginOrder} or cycles~\cite{goldbeter02}, while the attractors
of a signal transduction network correspond to steady state response(s) to the
presence of a given signal~\cite{tyson03}. In
Ref.~\cite{flyvbjerg1988esk,Drossel2005PRL,ShreimRBN2008} some exact results
were derived for RBNs with $K=1$. The critical case of $K=2$ was re-examined
numerically in Ref.~\cite{RBN-PL}, which reported power law behavior in the
distribution of transient times.  Ref.~\cite{DrosselAva} examined the
probability of returning to the same attractor after perturbing various number
of nodes. In Ref.~\cite{ShreimRBN2008} methods of complex network analysis
were able to distinguish the $K=2$ (critical) ensemble from the others using
measures of network heterogeneity in the SSNs. These network measures include
node degree (a local measure) and path diversity (a global measure) as well as
variations in the path diversity between different realizations in the
statistical ensemble.

Inspired by the idea of self-organized criticality~\cite{BakSandPile}, here we
start by using avalanches to probe the structure of the SSNs. Avalanches are
the responses of a system in an attractor state to small perturbations (a bit
flip of a single Boolean element). These eventually die out and the system
returns to an attractor state, which will be the same or a different
attractor.  By exact numeration, we find that the distribution of avalanche
times allows a clear distinction between the $K=2$ ensemble and other values
of $K$. While the avalanche and transient time distributions converge (except
for an overall normalization) at large times, significantly more avalanches of
short durations exist compared to transients.  This shows that attractors
preferentially cluster in configuration space for RBNs for all $K$ small
compared to the number of elements in the network, $N$.  This feature is also
found in the Boolean representation of the segment polarity gene network in
\emph{Drosophila}~\cite{albert2003dros}. This biological network has a
distribution of avalanche times closer to the $K=2$ ensemble than to other
values of $K$.

In order to clarify the differences between avalanches and transients at short
times we define a ``branching ratio'' to descibe how the average number of
dynamical states grows as a function of distance from the attractor. The
average branching ratio is given by the ratio of the probability distribution
for transient times at successive times, $P(T_{t}+1)/P(T_{t})$.  For $2\leq K
\ll N$ the quantity crosses unity at a certain distance from the attractor
where the SSN is the most ``bushy''.  The crossing time grows with system
size, $N$, indicating a new diverging time scale for RBNs.  This scale may
also separate the short time regime where the avalanche and transient times
differ from the long time regime where they converge.

\subsection{Outline}
In Section II, we review basic facts about RBNs and define relevant quantities
for our analysis.  Section III contains results from numerical simulations for
RBNs, while Section IV compares the behaviors found with that in the
\emph{Drosophila} segment polarity network.  We conclude with a summary in
Section V.

\section{Random Boolean Networks \label{rbns}}

An RBN consists of $N$ Boolean $(0,1)$ elements where the value of each
element evolves in discrete time according to a random Boolean function of $K$
distinct input arguments. Each of these arguments is chosen randomly from the
$N$ Boolean elements of the network. For each set of values of the arguments,
the Boolean function is chosen randomly to be $1$ with probability (bias) $p$
and $0$ with probability $1-p$. The inputs and functions assigned to each
element remain fixed for each realization. Ensemble averages are achieved here
by considering many different realizations of the Boolean network with $K$ and
$N$ constrained to certain values and for bias, $p =0.5$.

The state of the RBN is a bit-vector that specifies the value of each Boolean
element. Here we are exclusively concerned with the case of deterministic
dynamics achieved by updating all $N$ elements of a given RBN in parallel.
This uniquely maps each state to one successor state, known as its ``image''.
Consequently, the dynamics of an RBN can be visualized as a state space
network (SSN)~\cite{shreim07,ShreimRBN2008}. An SSN is made by connecting each
of the ${\cal N}=2^N$ dynamical states of an RBN to its image with a directed
link. The out-degree of each node in the SSN is the number of its images and
hence exactly one. The in-degree of a node, which is the number of pre-images
of the state, ranges from 0 to ${\cal N}$ in principle. In the context of
SSNs, the ``distance'' between two states or nodes in the network is the
length of the shortest path connecting them --- if such a path exists. This
directly implies that the distance between a state and its image is one.
Alternatively, one can use the Hamming distance (number of different bits
between two state vectors) as a metric. Whenever we do that, we call the set
of states ``configuration space''.  In general, there is no simple
relationship between these two measures of distance, {\it i.e.} between the
configuration space and the state space.

In the limit $K=N$, every Boolean element has the same set of inputs and its
dynamics is a random function of the entire state of the RBN. Since the
Boolean functions are chosen randomly this implies that the image for each
state coincides with the definition of a random map.  By construction, the
associated SSN of a random map is a random directed graph with a Poisson
distribution for its in-degree, with a mean of one while the out-degree of
each node is fixed to be one.

For finite $N$, any initial state eventually evolves to an attractor, which
may be a single state or a periodic cycle.  Depending on the specific RBN, an
arbitrary positive integer of different attractors can coexist within a single
realization of an SSN. Disconnected basins of attraction can occur -- each of
which consists of all states that evolve to the same attractor. Each state
that does not belong to an attractor is called a transient state and is
visited only once. Naturally, one can assign a quantity to
each attractor cycle which characterizes the probability $W$ that starting
from a random node in the SSN, the system is found in any particular state of
this attractor cycle after an arbitrarily long time (in the steady state). For
a state on the $i$th attractor cycle, $W_i$ is
\begin{eqnarray}
W_{i} = \frac{{B_i}}{{\cal N}{A_i}},
\label{attWeight}
\end{eqnarray}
where ${B_i}$ is the total number of states in the basin of attraction and
${A_i}$ is the length  of the attractor cycle.

We study the probability distribution of transient times and avalanche
durations for ensembles of RBNs with fixed $K$ and $N$. The transient time,
$T_t$, for a given initial state is defined as the number of time steps
required to reach an attractor. States constituting attractor cycles are
assigned a transient time $T_t=0$.  Here, the distribution of transient times
is obtained by considering all possible initial states (via exact enumeration)
over many independent realizations of the RBN.

Avalanches, on the other hand, are created by flipping a single Boolean
element ($0,1 \rightarrow 1,0$) of an attractor state. This definition
resembles avalanches in the ``Game of Life''~\cite{PerBakGoL,GoLBook}. An
avalanche continues until the system returns to a (possibly different)
attractor. Hence, the avalanche durations, $T_a$, are the transient times for
the collection of initial states generated by single bit flip perturbations to
attractors. For a given RBN, the distribution of these avalanche durations is
obtained from all possible avalanches created by every single Boolean element
flip of each attractor state.

Using exact enumeration avoids many potential biases typically encountered
when estimating properties of the dynamics as, for example, those related to
excluding very long transients due to computational constraints.  The
trade-off is that we only examine relatively small systems, $N \approx 25$, --
although the SSNs are large ${\cal O}(10^8)$~\footnote{Our results in this
  work are confined to RBNs with $N$ varying from $7$ to $23$ and random maps
  with $N$ varying from $7$ to $29$.}.

To make our results comparable to other studies that use random sampling
rather than exact enumeration~\cite{RBN-PL}, each avalanche is given a weight
$W_i$ defined in Eq.(\ref{attWeight}) and the weights are accumulated over
many realizations in order to obtain a probability distribution for avalanche
times in the RBN ensemble. We will discuss the effects of different weighting
schemes in a future publication~\cite{berdahl08b}.

Finally, we define the average branching ratio as the ratio of the number of
states (in the RBN ensemble) at distance $T+1$ from an attractor to the number
of states at distance $T$.  We study the finite size properties of this
quantity and find indications for a hitherto unknown, diverging time scale for
RBNs that cannot exist in the random map.

\section{Results for Random Boolean Networks}

\begin{figure}[htbp]
\begin{center}
\includegraphics*[width=\columnwidth]{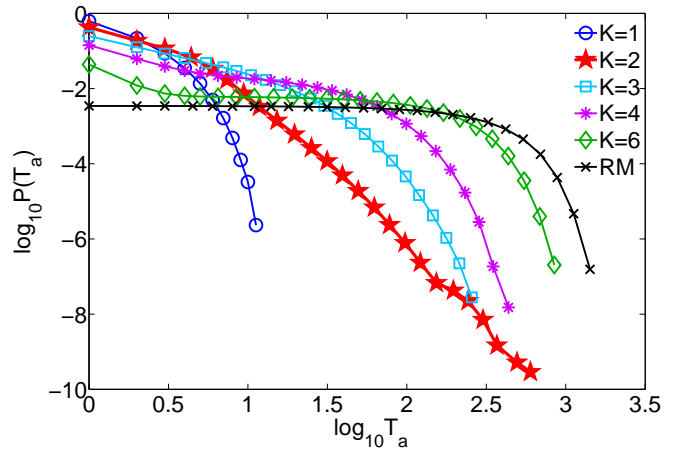}
\caption{\label{TallK-S17bigger}(Color online).  The probability
  distribution function (PDF) of avalanche times, $P(T_a)$, for
  various values of $K$ and the random map, all with $N=17$.  The
  $K=2$ curve stands out as a broad distribution without an apparent
  cut-off in our measurement window.  Numerical results are for $1.4\times
  10^6$ realizations for $K=1$, $4\times10^6$ for $K=2$, $9\times10^5$
  for $K=3$, $1\times10^6$ for $K=4$, $2.6\times10^5$ for $K=6$ and
  $1.3\times10^5$ for the random map.}
\end{center}
\end{figure}

\begin{figure}[htbp]
\begin{center}
\includegraphics*[width=\columnwidth]{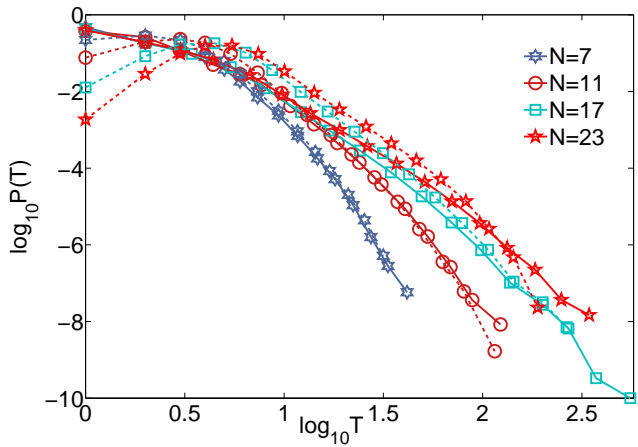}
\caption{\label{Time2Reach-AllAva-k=2loglog}(Color online).  Avalanche and
  transient times for $K=2$ RBNs with different $N$.  PDFs for avalanche times
  are indicated by solid lines. Dashed lines correspond to the PDFs,
  $P(T_{t})$, for transient times. These numerical results for avalanche
  (transient) times are based on the following number of realizations:
  $7.5\times10^6$ ($9\times 10^5$) for $N=7$, $10^7$ ($9\times 10^5$) for
  $N=11$, $4\times 10^6$ ($2.8\times 10^6$) for $N=17$, and
  $2.5\times 10^5$ ($2.6\times 10^4$) for $N=23$. Note the differences at short
  times while the long time behavior is statistically identical up to an
  overall normalization.}
\end{center}
\end{figure}

Fig.~\ref{TallK-S17bigger} shows the probability distribution function (PDF)
for avalanche durations, $P(T_a)$, for various values of $K$, as well as the
random map. For $K=1$, $P(T_a)$ is a narrow distribution. For $K \geq 3$, a
plateau in $P(T_a)$ appears that widens as $K$ increases.  For $K=3$ and $K=4$
the cut-off decays exponentially while for $K \geq 5$ and the random map, the
cut-off decays faster than exponentially with $T$. For $K=2$, $P(T_a)$ decays
slowly with no observed cut-off.  This is confirmed in
Fig.~\ref{Time2Reach-AllAva-k=2loglog}, which shows the dependence of $P(T_a)$
on $N$ for $K=2$. While the curves vary with $N$, no characteristic time scale
appears within our statistical resolution. This suggest that avalanche
durations are an indicator of criticality in RBNs.

\begin{figure}[htbp]
\begin{center}
\includegraphics*[width=\columnwidth]{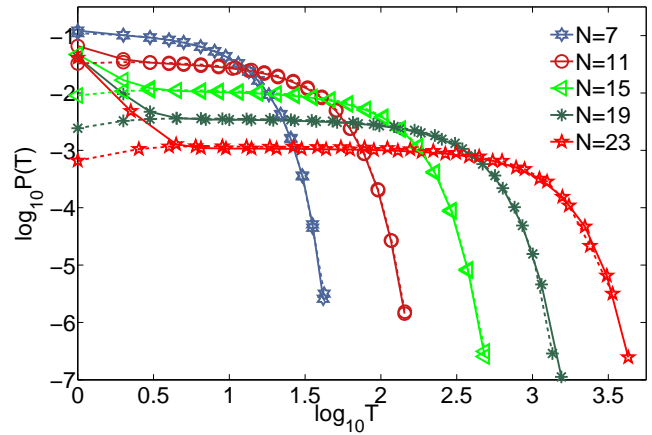}
\caption{\label{Time2Reach-AllAva-k=6loglog}(Color online).  Avalanche and
  transient times for $K=6$ RBNs. The PDFs for the avalanche (transient) times
  are indicated by solid (dashed) lines.  The results for avalanche
  (transient) times are based on the following number of realizations: $10^5$
  ($8\times 10^4$) for $N=7-15$, $2\times 10^4$ ($2\times 10^4$) for $N=19$,
  and $6\times 10^3$ (250) for $N=23$. Note that as for $K=2$, also for $K=6$
  there are differences at short times, while the long time behavior is
  statistically the same up to an overall normalization.}
\end{center}
\end{figure}

Fig.~\ref{Time2Reach-AllAva-k=2loglog} also compares $P(T_a)$ and $P(T_t)$ for
$K=2$. While $P(T_a)$ and $P(T_t)$ clearly differ for small arguments, the two
distributions become statistically indistinguishable, up to an overall
normalization, for larger arguments.  This is actually true for all values of
$K$ studied here --- see Fig.~\ref{Time2Reach-AllAva-k=6loglog} for the case
$K=6$. For larger $K$, the curves approach each other more quickly and for
$K=N$ the two PDFs are statistically and theoretically identical.  As
mentioned previously, the $K=N$ case corresponds to the random map, where no
correlation exists between the Boolean values of the elements in a state and
its image. As a result, flipping a single bit on an attractor state can put
the system into a state anywhere in the entire state space, and the
distribution of avalanche times, $P(T_a)$, for the random map is identical to
the distribution of the transient times, $P(T_t)$.

Ref.~\cite{RBN-PL} claimed that the PDF for \emph{transient times} for $K=2$
RBNs has a power-law tail for large $N$. Based on our results shown in Fig. 2
this also suggests that the distribution of avalanche times could have a
power-law tail for large $N$ reminiscent of self-organized critical
systems~\cite{bak}.  Note, however, that the purported ``exponent'' of the
power-law decay in Ref.~\cite{RBN-PL} varies systematically with system size
(as shown in Fig.~\ref{Time2Reach-AllAva-k=2loglog}) and that the behavior in
the limit of infinite system size remains unclear.  One possibility is that
(as discussed in detail later) the position of maximum in the transient time
distribution increases roughly with $N$ while the largest possible time is
bounded by the number of states ${\cal N}=2^N$.  This would give an
asymptotically uniform distribution at large times for large $N$.

\begin{figure}[htbp]
\begin{center}
\includegraphics*[width=\columnwidth]{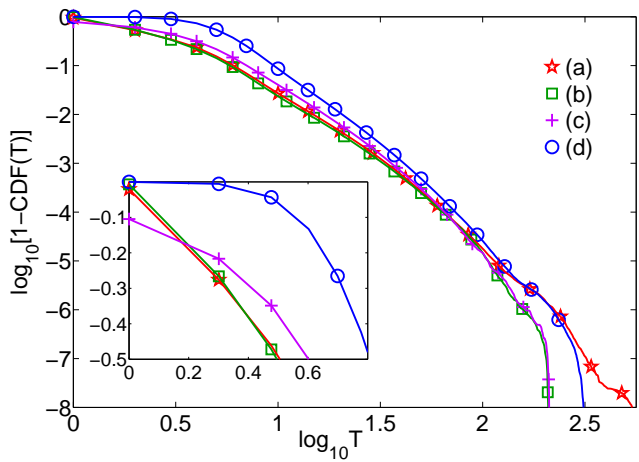}
\caption{\label{SameDiffCUMAK2}(Color online).  A comparison of the cumulative
  distribution functions (CDFs) in $K=2$ RBNs for: (a) the duration of all
  avalanches (b) the duration of avalanches that return to the same attractor
  (c) the duration of avalanches that lead to a different attractor (d) the
  transient times. The inset shows a linear magnification of the upper left
  section of the original plot. Curves are based on the following numbers of
  realizations of RBNs with $N=17$: (a) on $4\times 10^6$, (b) and (c) on
  $7\times10^5$ and (d) on $2.8\times 10^6$.}
\end{center}
\end{figure}

The aforementioned differences between the PDFs for avalanche times and
transient times are characterized by $P(T_a) > P(T_t)$ for small arguments for
all studied values of $N$ with $K<N$. Both the transient time and avalanche
duration are the time it takes for the system to reach an attractor from an
initial state. While for the transient times the initial state is chosen
randomly, for avalanche times the initial state is chosen by making a
perturbation to an attractor state. We conclude that a state generated by a
single flip perturbation to an attractor state tends to be closer to an
attractor state in the SSN than a randomly chosen state. It also has a
statistically significant chance to be, itself, an attractor state on a
different attractor.

Fig.~\ref{SameDiffCUMAK2} shows the cumulative distribution (CDF) of avalanche
durations for all perturbations, for perturbations that remain in the same
basin of attraction and for those that lead to a different basin of attraction
and compares them with the CDF of transient times, for $K=2$. A single element
perturbation leads to a nearest neighbor in configuration space.
Fig.~\ref{SameDiffCUMAK2} shows that these neighbors in configuration space
are more likely to be close to a (possibly different) attractor in the SSN
than randomly selected states.  Fig.~\ref{SameDiffCUMAK2} also indicates that
there is a non-negligible probability that avalanches that go to different
attractors have duration zero.  This is indicated by the fact that the curve
$c$ does not go through the point (0,0). It means that one bit flip is
sufficient to put an attractor state directly onto a different attractor. This
suggests in particular the existence of regions in \emph{configuration space},
where the attractors are more likely to live: Different attractors in the SSN
are often close in configuration space.

Qualitatively similar results hold for all $K \ll N$. However, independent of
$N$, the fraction of avalanches that return to the same attractor
monotonically decreases as $K$ increases, starting at $95\%$ for $K=1$ and
asymptotically converging to 2/3 for the random map for large
$N$~\cite{DrosselAva}. Moreover, the PDF for the duration of avalanches that
lead to different attractors as well as the corresponding PDF for avalanches
leading to the same attractors approach the PDF for transient times as $K$
approaches $N$.

\begin{figure}[htbp]
  \begin{minipage}[c]{0.7\columnwidth}
    \centering
    \includegraphics*[width=0.7\columnwidth]{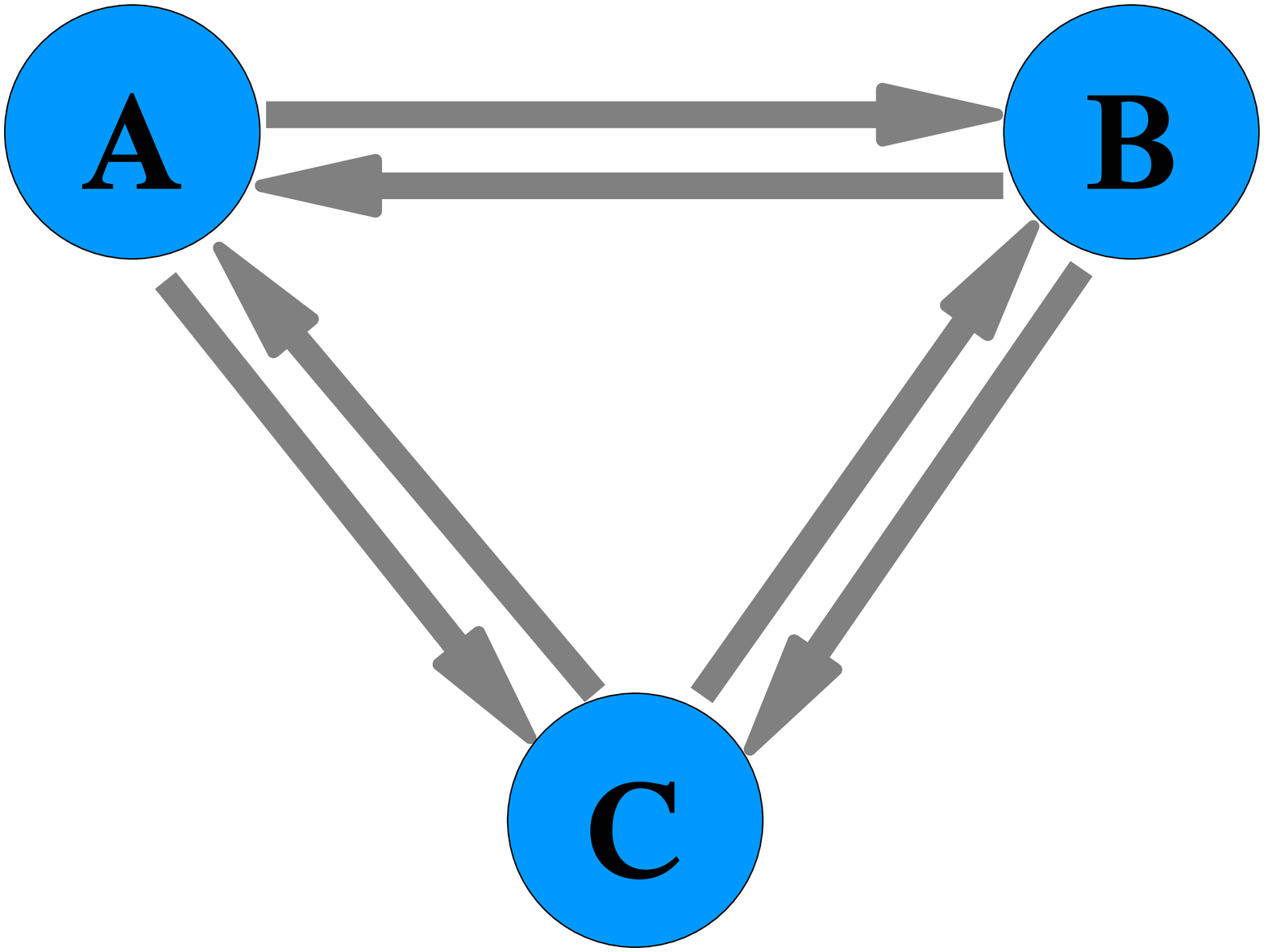}
  \end{minipage}
  \vskip 1cm
  \begin{minipage}[c]{0.7\columnwidth}
    \centering
    \begin{tabular}{l l|l l l l|l l l l|l }

      B & C & A & ~\ ~\ ~\ & A & C & B & ~\ ~\ ~\ & A & B & C  \\
      \cline{1-3}
      \cline{5-7}
      \cline{9-11}
      0 & 0  & 0 & & 0 & 0  & 0 & & 0 & 0  & 1\\
      0 & 1  & 0 & & 0 & 1  & 0 & & 0 & 1  & 1\\
      1 & 0  & 1 & & 1 & 0  & 1 & & 1 & 0  & 1\\
      1 & 1  & 1 & & 1 & 1  & 1 & & 1 & 1  & 1\\
    \end{tabular}
  \end{minipage}

  \caption{\label{Illustration} The network structure for an illustrative
    example of an RBN with $K=2$.  Elements $A$ and $B$ copy each other at
    each time step while element $C$ is forced to 1.}
\end{figure}

To see why the attractors are clustered in configuration space, we consider
the notion of relevant and irrelevant components (for a thorough discussion
see e.g.~\cite{Drossel2006,socolar2003soa}).  The elements of an RBN can be
divided into relevant and irrelevant elements.  The relevant elements are
clustered into what are called relevant components.  These components
determine the number and lengths of
attractors~\cite{Drossel2006,socolar2003soa}. The irrelevant elements are
those whose information is lost in the dynamics. As an example, consider the
simple network in Fig.~\ref{Illustration}.  The network is made of the
relevant component made of elements $A$ and $B$ and the irrelevant component
element $C$.  $A$ and $B$ form an information conserving loop by copying each
other, while $C$ is forced to $1$ regardless of the state of its inputs.
Moreover, $C$ does not influence the state of $A$ and $B$.  Consider the
attractor state $[\sigma(A), \sigma(B), \sigma(C)] = [0, 0, 1]$, where
$\sigma(X)$ is the value of the element $X$. If we flip element $C$ the system
will return to the same attractor $[0, 0, 1]$ once the information injected
into the irrelevant component is lost.  Alternately we could flip element $A$
which is part of the relevant component.  This would lead to state $[1, 0,
1]$, which is a state on a period two attractor cycle composed of $[1, 0, 1]$
and $[0, 1, 1]$.  This perturbation leads directly to a different attractor.
In general we can replace the element $C$ by a set of elements $\{ C_{i} \}$
which are neither influenced by nor influence $A$ and $B$.  This allows us to
the see the clustering of attractor states; the attractor states: $[0, 0,
\sigma(\{ C_{i} \})]$, $[0, 1, \sigma( \{C_{i} \})]$, $[1, 0, \sigma( \{C_{i}
\})]$ and $[1, 1, \sigma( \{C_{i} \})]$ are clustered into four neighboring
sites in configuration space.

In general flipping the state of an irrelevant element on an attractor will be
forgotten and the dynamics eventually leads to the same attractor.  However,
by flipping the state of a relevant element, the system can reach a different
basin of attraction.

\begin{figure}[htbp]
\begin{center}
\includegraphics*[width=\columnwidth]{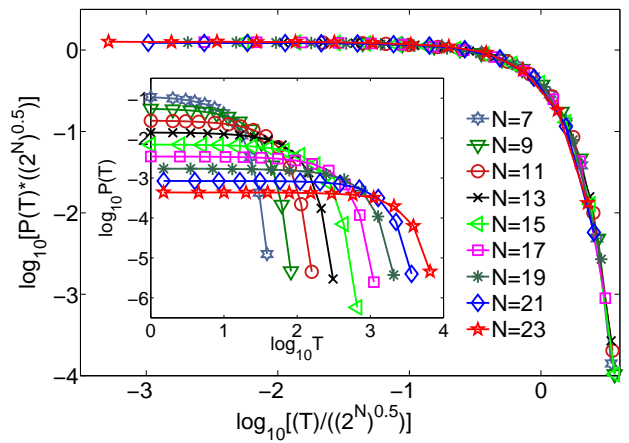}
\caption{\label{TryInsetRM}(Color online).  The PDF $P(T_{t})$ for the
  duration of transients in random maps (used to simulate the SSNs of $K=N$
  RBNs) for various $N$ collapsed using Eq. (\ref{trappx}).  The inset shows
  the un-scaled PDFs. The numerical results for $N=7-17$ are based on
  $1.4\times 10^5$ realizations of the random map, $N=19$ on $3.4\times 10^3$
  and $N=21-23$ on 500.}
\end{center}
\end{figure}

In the chaotic phase, $K>2$, the distribution of transient times can be
estimated from the joint probability distribution for transient times and
attractor lengths derived in \cite{bastollaparisi} using an annealed
approximation,
\begin{eqnarray}
\label{trappx}
P[T_t] &\approx& \frac{1}{e^{\alpha N}}\sum_{A=1}^{{2^N}-T_t}{\rm exp}
\left[-\frac{1}{2} \left(\frac{(T_t+A)}{e^{\alpha N/2}}\right )^2\right]\nonumber\\ 
&\approx& \frac{1}{e^{\alpha N/2}} \int_{(T_t+1)/(e^{\alpha N/2})}^{\infty}dx e^{-x^{2}/2},
\end{eqnarray}
for large $N$ and $T_t$. In general, $\alpha$ depends on $K$. It was shown in
Ref.~\cite{bastollaparisi} that $\alpha$ vanishes for $K=2$, and approaches
the value ${\log 2}\approx 0.69$ for the random map.  Bastolla and Parisi also
pointed out that, while the analytical expressions for the transient time
distribution may not be good approximations for small $K$, the time scale
$e^{\alpha N/2}$ derived using the annealed approximation (see
Eq.~(\ref{trappx})) is in good agreement with their simulations for $K=3$.

\begin{figure}[htbp]
\begin{center}
\includegraphics*[width=\columnwidth]{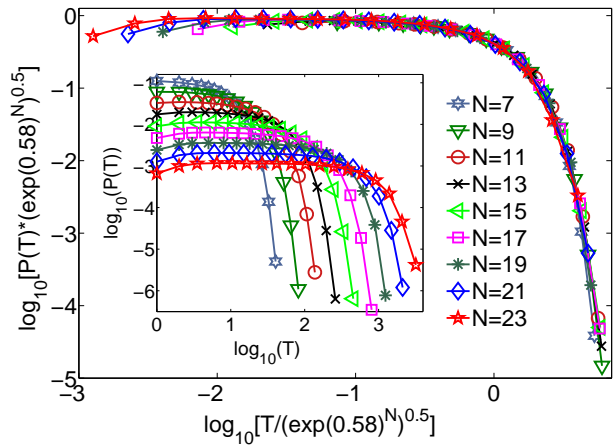}
\caption{\label{InsetK6}(Color online).  The PDF $P(T_{t})$ for the duration
  of transient times in $K=6$ RBNs collapsed using Eq.  (\ref{trappx}). The
  inset shows the un-scaled PDFs. The numerical results for $N=7-15$ are based
  on $8\times 10^4$ realizations of the random map, $N=17$ on $2.5\times
  10^5$, $N=19-21$ on $2\times 10^3$ and $N=23$ on $250$.}
\end{center}
\end{figure}

This last result is confirmed by our numerical simulations for various values
of $K$. Fig.~\ref{TryInsetRM} shows that $P(T_t)$ for the random map and
various $N$ indeed exhibits finite size scaling with characteristic time
$2^{N/2}$. For $K=6$ we also find a reasonable data collapse, shown in
Fig.~\ref{InsetK6} using the value $\alpha=0.58$ obtained in
\cite{bastollaparisi}. Yet, the quality of the data collapse continues to
deteriorate as $K$ decreases to $3$.

\begin{figure}[htbp]
\begin{center}
\includegraphics*[width=\columnwidth]{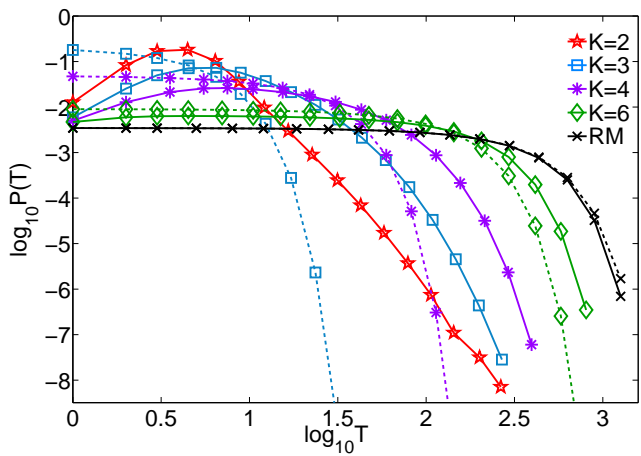}
\caption{\label{tranfits}(Color online).  The PDF, $P(T_{t})$, for transient
  lengths in the SSNs of RBNs for various values of $K$ and the random map,
  $N=17$. Each (excluding $K=2$) is accompanied by a corresponding theoretical
  curve based on Eq. (\ref{trappx}) shown with dashed line. Numerical results
  are for $2.8\times 10^6$ realizations for $K=2$, $8\times10^5$ for $K=3$,
  $9\times10^5$ for $K=4$, $2.5\times10^5$ for $K=6$, and $1\times10^4$ for
  the random map. Based on \cite{bastollaparisi}, we used the values $\alpha =
  0.20, 0.38, 0.58$ and $\log 2$ for $K = 3, 4, 6$ and the random map
  respectively.}
\end{center}
\end{figure}

Fig.~\ref{tranfits} shows the distribution of transient times along with the
theoretical predictions given by Eq.~(\ref{trappx}). For the random map, the
prediction is exact and the agreement between data and theory is excellent.
For $K=6$ the theoretical predictions do not match the data for small $T_t$.
The theoretical curve in Eq.(\ref{trappx}) decreases monotonically with $T$
for all $K>2$. However, the actual distribution increases initially with $T_t$
for all $K<N$ considered. This increase becomes more prominent for smaller
$K$.

The number of states in the ensemble that are at a distance $T_t$ from an
attractor is $2^N P(T_t)$. Considering the arboreal structure of the SSN, this
number corresponds to the number of states in the $T_t$-th shell. An increase
of $P(T_t)$ for small $T_t$ then means that the shells become more populated
with distance from an attractor. The monotonic decrease of Eq.~(\ref{trappx})
indicates that the $T_t=0$ (attractor) shell should be the most populated,
which we do observe for the random map. The maxima observed in
Fig.~\ref{tranfits} for RBNs with $K<N$ indicate the existence of a new
characteristic distance for RBNs with $K<N$ that does not appear in the random
map.

\begin{figure}[htbp]
\begin{center}
\includegraphics*[width=\columnwidth]{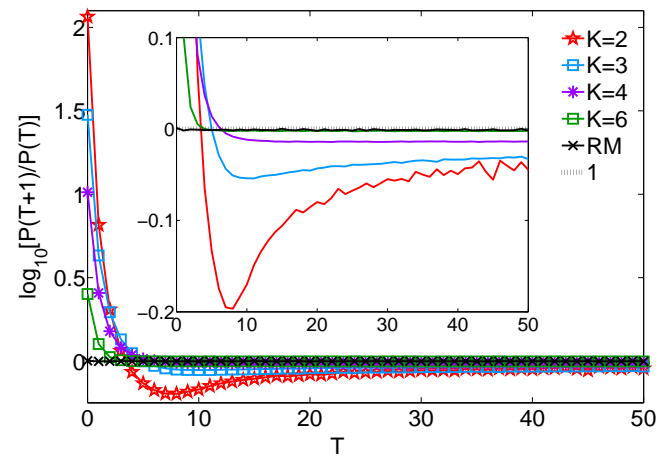}
\caption{\label{branching}(Color online). The mean branching ratio,
  $P(T_t+1)/P(T_t)$, for the arboreal structure of the SSNs for RBNs for
  various values of $K$ and the random map with $N=17$. The inset shows a
  magnification of the same curves. The mean branching ratio of $1$ --- which
  is the mean degree of the SSN --- is shown for comparison.  The results are
  for $2.8\times 10^6$ realizations for $K=2$, $8\times 10^5$ for $K=3$,
  $9\times 10^5$ for $K=4$, $2.5 \times 10^5$ for $K=6$, and $1.3 \times 10^5$
  for the random map.}
\end{center}
\end{figure}

The ratio of the sizes of consecutive shells are shown in
Fig.~\ref{branching}. A state in the $(T_t+1)$-th shell can be seen as
branching from its image in the $T_t$-th shell. If each basin of the SSN is
viewed as a tree with the attractor at its root, Fig.~\ref{branching} gives a
mean branching ratio from the $T_t$-th shell to the $(T_t+1)$-th shell for the
whole ensemble. Note that this is \emph{not} the mean of the branching ratios
for each attractor basin.  For $K<N$, Fig.~\ref{branching} shows that for
small $T_t$ the shells first grow at a rate that slows down with distance from
the attractors. The growth eventually stops and the shells become smaller and
smaller for increasing $T_t$ when it is larger than a certain value that
depends on $K$ (and on $N$). This turn around happens when the ratio drops
below unity.  For the random map, Eq.~(\ref{trappx}) implies that shells
should become smaller for all $T_t$, albeit very slowly. This can be seen in
Fig.~\ref{branching} in which the random map curve almost coincides with
unity.  The $K=6$ curve crosses unity and quickly merges with the random map
curve, while the $K = 4, 3$ and $2$ curves make a discernible dip below the
$P(T_t+1)/P(T_t) = 1$ line, the extent of the dip depending on $K$. Indeed it
appears that the behavior of the branching ratio for $K=2$ RBNs shows more
variation than other values of $K$.

\begin{figure}[htbp]
\begin{center}
\includegraphics*[width=\columnwidth]{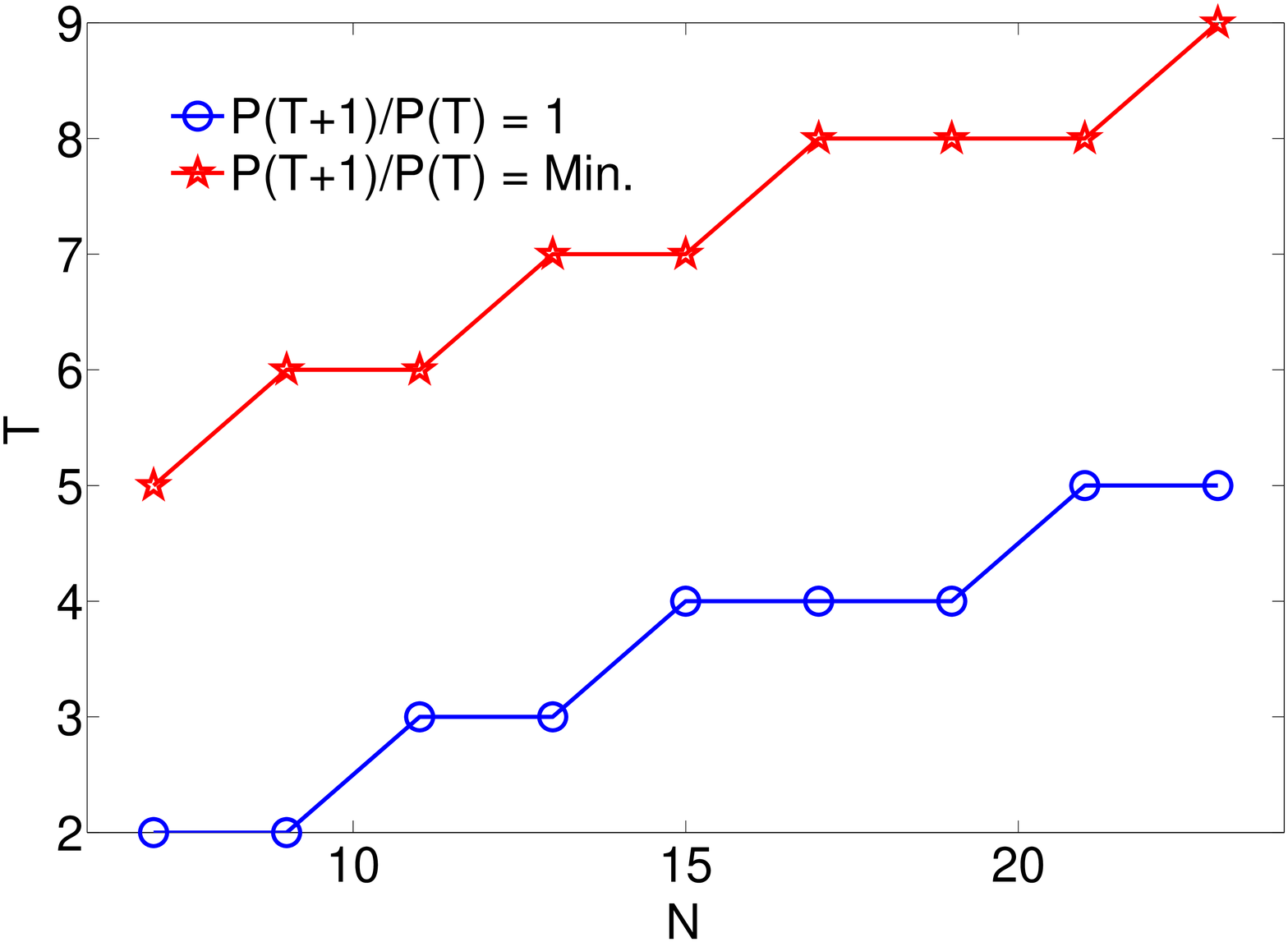}
\caption{\label{CrossnDip}(Color online). The value of $T_t$ where the mean
  branching ratio (Fig.~\ref{branching}) is minimal and the value where
  $P(T_t+1)/P(T_t) = 1$ for various sizes of RBNs with $K=2$.}
\end{center}
\end{figure}

Both the value of $T_t$ where the branching ratio is minimal and the value
where $P(T_t+1)/P(T_t) = 1$ increase with increasing $N$ as shown for $K=2$ in
Fig.~\ref{CrossnDip}.  The roughly linear increase also appears to hold for
$2<K<N$.  If the growth that we find in Fig.~\ref{CrossnDip} can be
extrapolated to large system sizes, our results indicate the existence of a
new diverging time scale for RBNs with $2<K \ll N$ that is unrelated to the
cut-off for large arguments given by Eq.~(\ref{trappx}).

\section{Results for the segment polarity network of \emph{Drosophila
    melanogaster}}

To compare our results for RBNs with a biological signaling network, we
consider the segment polarity network of fruit flies~\cite{albert2003dros}.
This Boolean model of the gene and protein interactions involved in embryonic
pattern formation in the fruit fly \emph{Drosophila melanogaster} is a
particularly well-documented and successful application. In particular, it
reproduces many experimentally observed features, including knock out results,
indicating that knowledge of the kinetic details are not necessary.

The model presented in~\cite{albert2003dros} considers gene expression
patterns in four adjacent cells that form a \emph{Drosophila} parasegment.  Each cell
is modelled by a network of fifteen Boolean elements. Each element represents
either an mRNA species or a protein species in the cell.  The state of each
element is one if the corresponding species is expressed, otherwise it is
zero. Each element in a cell can interact with elements within the same cell
as well as elements in the neighboring cells. All interactions are modelled by
Boolean functions. Of the 15 species, the state of protein ``sloppy-paired''
(SLP) in each of the 4 cells is fixed to its biologically relevant value. The
resulting state space contains 6 fixed point attractors.

Here we study the dynamics of only one cell, for two reasons.  First, the
complete network, which consists of 60 elements, is computationally
untractable using exact enumeration. Second, a single cell represents the
building block of the entire organism.  Understanding the dynamics of a single
cell is important to understand the whole network. Robust dynamics of the
individual cells that form the segment has been argued to underlie the
observed robustness of expression patterns of the whole
segment~\cite{ingolia2004}.A differential equation model for individual cells
based on the original work in~\cite{vondassow2000} was used to show this
relation between the dynamics of the individual cells and the whole segment.
In the same vein, here we study the state space of the Boolean network for an
individual cell. As a result, our individual cell network cannot be directly
compared with the parasegment network studied in~\cite{albert2003dros}.

In the Boolean network model of~\cite{albert2003dros}, cells interact via
three elements, the hedgehog mRNA (hh), the hedgehog protein (HH), and the
wingless protein (WG). To reduce the model to one individual cell, we replace
the values of these three elements in the cell's neighbourhood by three new
variables, hhN, HHN and WGN. Each of these three elements evolves by copying
its state.  Furthermore SLP also copies its own state.  Therefore, the values
of the states of these four elements are fixed by the initial conditions and
do not evolve dynamically.  Note that this ensures the existence of \emph{at
  least} $16$ different attractors, each corresponds to a unique combination
of the input elements.

The polarity network studied here contains $15 + 3 = 18$ elements in total.
The connectivity, $K$, of the nodes in the polarity network is distributed as
follows: $9$ nodes have $K=1$, $4$ have $K=2$, $4$ have $K=3$ and only one
node has $K=4$.  The mean connectivity in the network is thus $1.83$. The
Boolean rules (functions) of the nodes in the network are given in
Table~\ref{TableBooleanFun}. The average bias, $p$, in the network is $0.42$.
The internal homogeneity, which measures the average percentage of the most
abundant outcome (zero or one) in each Boolean function, is $0.625$.

\begin{table}[!ht]
\begin{center}
\begin{tabular}{ l||l }

Node           & Boolean function (input/output relation)\\
\hline
\hline
wg              & (CIA AND SLP AND NOT CIR) \\
                &  OR (wg AND (CIA OR SLP) AND NOT CIR)\\
\hline
WG              &  wg\\
\hline
en             & WGN AND NOT SLP\\
\hline
EN              & en\\
\hline
hh              & EN AND NOT CIR\\
\hline
HH              & hh\\
\hline
ptc             & CIA AND NOT EN AND NOT CIR\\
\hline
PTC             & ptc OR (PTC AND NOT HHN)\\
\hline
PH              & PTC AND HHN\\
\hline
SMO             & NOT PTC OR HHN\\
\hline
ci              & NOT EN\\
\hline
CI              & ci\\
\hline
CIA             & CI AND (SMO OR hhN)\\
\hline
CIR             & CI AND NOT SMO AND NOT hhN\\
\hline
SLP             & SLP (\emph{input})\\
\hline
hhN             & hhN (\emph{input})\\
\hline
HHN             & HHN (\emph{input})\\
\hline
WGN             & WGN (\emph{input})\\
\hline
\end{tabular}
\caption{\label{TableBooleanFun} The Boolean function of each node in the
\emph{Drosophila} segment polarity network. Upper case names are reserved for
proteins and the lower case names are for mRNA. The hhN, HHN and WGN nodes
refer to the state of hh, HH and WG in the neighboring cell.}
\end{center}
\end{table}

\begin{figure}[htbp]
\begin{center}
\includegraphics*[width=\columnwidth]{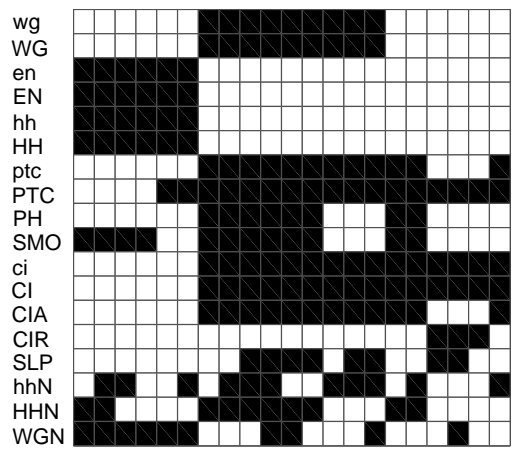}
\caption{\label{Attractors} The attractor states of the \emph{Drosophila}
  polarity network. Each column represent an attractor, the dark and white
  square represent 1 and 0 respectively. The order of the attractors was
  chosen to show attractor clustering in configuration space.}
\end{center}
\end{figure}

We find that the individual cell state space has $21$ different attractors,
each of which is a fixed point.  The states of the 21 attractors are shown in
Fig.~\ref{Attractors}. Each of $21$ columns in the figure represents an
attractor. Each of $18$ rows represent the value of the Boolean state of a
particular element across different attractors. The black squares depict ones
while white squares depict zeros.

Fig.~\ref{Attractors} also illustrates the clustering of attractors in
configuration space.  Attractors in the figure are arranged in clusters where
neighboring attractors in each cluster are one bit flip away. Note that the
bit flips that immediately lead to a different attractor are all flips of the
relevant component. In this context, there are at least 4 relevant components
in the network~\footnote{Note that the existence of $21$ attractors instead of
  $16$ is an indication of the existence of further relevant components. For
  instance, one can notice that nodes wg and PTC copy themselves under certain
  configurations of their input elements, thus becoming relevant. However, bit
  flips to these elements doesn't lead directly to another attractor.}.  SLP,
hhN, HHN, and WGN all copy their own state at each time step, therefore each
is a relevant component.

We note that the attractors in Fig.~\ref{Attractors} capture the cell states
observed in \emph{Drosophila} segments. For example, none of the cells in the
\emph{Drosophila} segment co-express genes wg and en~\cite{sanson2001gpf}.
Lack of wg-en co-expression is also seen in the attractors in
Fig.~\ref{Attractors}.  Our results suggest that the observed gene expression
phenotype of the segment polarity network stems from the constrained dynamics
in a single cell, rather than the complex interactions between the cells. The
same has been argued previously in~\cite{ingolia2004} using a differential
equation model for a single cell. The biological significance of these results
and their impact on gene expression robustness will be expounded in a future
publication~\cite{SoodDros2008}.

\begin{figure}[htbp]
\begin{center}
\includegraphics*[width=\columnwidth]{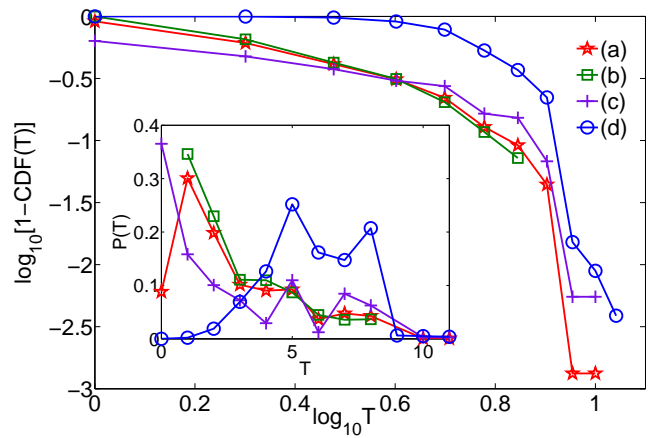}
\caption{\label{Polarity}(Color online).  Comparison of CDFs from the
  segment polarity network of \emph{Drosophila} ($N=18$) for: (a) the
  duration of all avalanches (b) the duration of avalanches that
  return to the same attractor (c) the duration of avalanches that
  lead to a different attractor (d) the transient times. The inset
  shows the PDFs for the same data.}
\end{center}
\end{figure}

Avalanches and transients for the segment polarity network are defined in the
same way as for RBNs --- see section~\ref{rbns}. Fig.~\ref{Polarity} shows the
distributions of avalanche and transient times for the segment polarity
network in Table~\ref{TableBooleanFun}.  As for RBNs (see, for example,
Fig.~\ref{SameDiffCUMAK2}) there are more short avalanches than short
transients indicating that attractors cluster in configuration space. This
holds for the specific avalanches that return to the same attractor as well as
those that go to a different attractor.  Interestingly, in the latter case
there is a significant probability that an avalanche will immediately reach
another attractor --- similar to the case of RBNs with $K \ll N$.

\begin{figure}[htbp]
\begin{center}
\includegraphics*[width=\columnwidth]{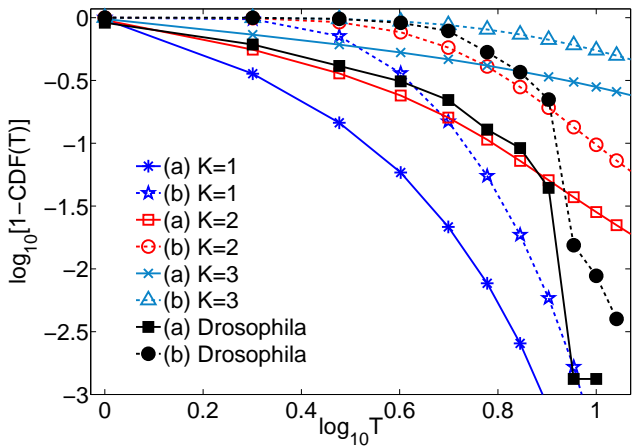}
\caption{\label{Polarity_cdf}(Color online). Comparison of CDFs from the
  segment polarity network of \emph{Drosophila} ($N=18$) with those for RBNs
  with $N=18$ and $K = 1, 2$ and $3$. For each case we show the results for
  (a) the duration of all avalanches and (b) the transient times. Note that
  the data for \emph{Drosophila} correspond to a \emph{single} Boolean network
  while the data for the RBNs correspond to ensemble averages.}
\end{center}
\end{figure}

The qualitative similarities between RBNs and the \emph{Drosophila} polarity network
are also demonstrated in Fig.~\ref{Polarity_cdf} where the distributions for
RBNs with $K=1,2,3$ are shown for comparison.  The large scatter, due to the
fact that we are examining only one network rather than an ensemble, prevents
any precise quantitative comparison between the segment polarity network and
RBNs. It is interesting to note, though, that the distributions of avalanche
durations for the polarity network have a tendency to be more similar to RBNs
with $K=2$ at small values of $T$ than any other $K$.

\section{Conclusions}

We have studied avalanches on Random Boolean Networks and on the segment
polarity network of \emph{Drosophila} by performing bit-flip perturbation to
attractor states and waiting for the dynamics to relax back to an attractor.
In both cases (assuming $K \ll N$ for RBNs), there are more avalanches of
short duration than what would be expected based on the distribution of
transients times. This indicates that attractors tend to cluster in
configuration space.

For the random map, which corresponds to RBNs with $K=N$, the distribution of
avalanche durations, $P(T_a)$, is identical to the distribution of transient
times, $P(T_t)$. As analytically confirmed, the distribution tends to form a
plateau with a sharp finite-size cut-off. For $2 < K < N$, deviations from
this behavior occur for small arguments such that the plateau eventually
disappears as $K \rightarrow 2$. In particular, the deviations for small
arguments are different for $P(T_a)$ and $P(T_t)$. The critical case $K=2$
exhibits different behavior. In this case, both distributions are broad and
neither the plateau nor the cut-off are observed. Their scale-free appearance
suggests that the distribution of avalanche durations as well as the
distribution of transient times can be used as possibly distinct indicators
for the criticality in discrete, deterministic dynamical systems.  Indeed we
find that the avalanche durations for the \emph{Drosophila} segment polarity
network follows more closely the $K=2$ behavior (compared to other values of
$K$) while no clear statement can be made about transient times for the
\emph{Drosophila} network we analyzed.

The similarity of $P(T_a)$ and $P(T_t)$ for large arguments and $K>2$
indicates that initial avalanche states that are far away from an attractor
are independent of the states on the attractors (this is true for all
arguments in the case of the random map). The arboreal structure of the SSN,
the mean branching of which was presented in Fig~\ref{branching}, also seems
to become more similar to that of the random map for large $K$.

However, the branching ratio is a non-monotonic function of $T_t$, crossing
unity and then passing through a minimum before it starts to resemble the
random-map behavior for larger $T_t$.  Along with our results for the
differences between $P(T_a)$ and $P(T_t)$ for small arguments, this indicates
that while states lying closer to the attractors in the SSN are correlated
with the attractor states, this correlation decays as one moves away from the
attractors on the SSN. The time scale at which this correlation in the
arboreal structure of the SSN decays, or in which the growth rate of the tree
structure in state space changes from increasing to decreasing may be a new
characteristic scale for RBNs. It appears to diverge roughly linearly with the
system size $N$.  This time scale does not appear in the random map and is not
related to that previously studied in \cite{bastollaparisi}, which grows
exponentially with the system size $N$.  Beyond that scale, the SSN of an RBN
in the \emph{non-frozen} regime may correspond to a random map, while the
regions around attractors are distorted by their presence.

\begin{acknowledgments}
  We would like to thank Peter Grassberger for useful comments. This work was
  partially supported by NSERC.

\end{acknowledgments}

\bibliography{References.bib}

\end{document}